\documentclass[preprint]{aastex}
\pdfoutput=1
\usepackage{times}
\usepackage{wasysym} % symbols
\usepackage{caption} %subfigures
\usepackage{subcaption} %subfigures
\usepackage{hyperref} % internal links
\usepackage{natbib} % bibliography
\usepackage{multirow}

\usepackage{graphicx}
\let\oldtabular\tabular
\renewcommand{\tabular}{\scriptsize\oldtabular}

\usepackage[normalem]{ulem}

\begin{document}
\title{The Stability of F-star Brightness on Century Timescales}
\author{Michael B.\ Lund\altaffilmark{1},
Joshua Pepper\altaffilmark{2,1},
Keivan G.\ Stassun\altaffilmark{1,3},
Michael Hippke\altaffilmark{4},
Daniel Angerhausen\altaffilmark{5}
}
\altaffiltext{1}{Department of Physics and Astronomy, Vanderbilt University, Nashville, TN 37235, USA; \\ \url{michael.b.lund@vanderbilt.edu}}
\altaffiltext{2}{Department of Physics, Lehigh University, Bethlehem, PA 18015, USA}
\altaffiltext{3}{Department of Physics, Fisk University, Nashville, TN 37208, USA}
\altaffiltext{4}{Institute for Data Analysis, Luiter Stra{\ss}e 21b, 47506 Neukirchen-Vluyn, Germany}
\altaffiltext{5}{USRA NASA Postdoctoral Program Fellow, NASA Goddard Space Flight Center, Exoplanets \& Stellar Astrophysics Laboratory, Code 667, Greenbelt, MD 20771, USA}

\captionsetup[table]{labelsep=space}
\captionsetup[figure]{labelsep=space}

\begin{abstract}
The century-long photometric record of the DASCH project provides a unique window into the variability of stars normally considered to be photometrically inactive.  In this paper, we look for long-term trends in the brightness of F stars, with particular attention to KIC 8462852, an F3 main sequence star that has been identified as significant short-term variability according to \emph{Kepler} observations.  Although a simple search for variability suggests long-term dimming of a number of F stars, we find that such trends are artifacts of the "Menzel Gap" in the DASCH data.  That includes the behavior of KIC 8462852, which we believe is consistent with constant flux over the full duration of observations.  We do, however, present a selection of F stars that do have significant photometric trends, even after systematics are taken into account.
%The century-long DASCH light curves provide the opportunity to look for long-term trends in the brightness of F stars. We give particular attention to KIC 8462852, an F3 main sequence star that has been identified as significant short-term variability according to \emph{Kepler} observations. Fitting a single line to the century-long light curve suggests that there is a significant long-term variation as well, however when we analyze a large number of F-stars, we find that significant slopes appear to be very common. This would be inconsistent with the distribution of slopes that would be expected simply from statistical noise if F stars were of constant flux. When all the light curves are split during the "Menzel Gap", however, we find that there's a significant reduction in high-significance slopes, including that KIC 8462852 is consistent with constant flux. We also present a selection of F stars that still have significant slopes, even after this issue is taken into account.
\end{abstract}

\section{Introduction}
KIC 8462852 (TYC 3162-665-1, and eponymously nicknamed ``Tabby's Star") is an F3 V/IV star in the \emph{Kepler} field. It was discovered to undergo aperiodic, irregular dimming events, with reductions in flux as large as 20\% \citep{Boyajian2016}. The star  does not have any apparent infrared excess that would indicate the presence of a large amount of circumstellar dust \citep{Marengo2015, Thompson2015c}. Thus, one explanation that has been put forward  is a family of comets orbiting the star \citep{Bodman2016}. More exotic solutions have also been proposed, such as that KIC 8462852 is host to a Dyson swarm \citep{Wright2015a}, and SETI observations have produced null results for signals in optical and radio wavelengths \citep{Schuetz2015, Harp2015}.

The time baseline provided by \emph{Kepler} was expanded greatly by the incorporation of DASCH (Digital Access to a Sky Century@Harvard) archival data over the past century in an initial analysis carried out by \citet{Schaefer2016}. They provided comparisons between the light curve of KIC 8462852 versus those of two similar stars in close proximity to KIC 8462852 in the photographic plates. \citet{Schaefer2016} finds that KIC 8462852 has steadily dimmed at a rate of 0.165 mag/century over the entire DASCH light curve, which could favor alternatives to the comets hypothesis. 
In contrast, a follow-up analysis by \citet{Hippke2016a}, which used a larger sample of $\sim$50 DASCH comparison stars, finds that KIC 8462852 is not unique and that many DASCH stars show apparent long-term trends in their light curves. Those authors suggest that many of the stars experience structural breaks in their light curves, likely due to systematics, and are better modeled using two lines with a discontinuity at some year.
In particular, the so-called ``Menzel gap" in the 1950s and 1960s during which few observations were obtained, may coincide with a systematic offset in the full set of DASCH light curves that is responsible for the apparent long-term trends in the light curves of KIC 8462852 and many other stars.

In this paper, we aim to more comprehensively assess the stability of brightness on century timescales for a large sample of F stars in order to understand the phenomenon of long-timescale variations generally as well as to specifically fully assess the DASCH photometry for KIC 8462852 to determine the significance of any long-term trends. In Section 2, we present the DASCH photometry and the large set of comparison stars we have identified. In Section 3, we examine the long-term trend of KIC 8462852 as well as several hundred other comparable stars to establish the degree to which KIC 8462852 is typical of DASCH stars in terms of the apparent long-timescale trend in its light curve. We also examine if a systematic break at a single year can reduce the number of stars whose light curves appear to have significant long-term trends. Finally, in Section 4, we highlight a sample of stars that may have real long-term trends in their light curves after the likelihood of such trends have been ruled out for KIC 8462852 and for most other comparable stars. We conclude in Section 5 with a summary of our conclusions.

\section{Data}
\subsection{Digital Access to a Sky Century@Harvard (DASCH) project Light Curves}\label{sec:DASCH}
The Digital Access to a Sky Century@Harvard (DASCH) project is on ongoing project to digitize $\sim$ 500,000 glass photographic plates at Harvard College Observatory from observations with several telescopes from 1880 to 1985. These plates are scanned, and a data reduction pipeline is used to identify sources, calibrate each plate, generate light curves, and flag any potentially bad data points \citep{Laycock2010}. To calculate magnitudes, source stars are compared to catalog stars, and DASCH has used three different stellar catalogs for this purpose. The initial photometry was done by using the Hubble Guide Star catalogue (GSC2.3) B-band magnitudes to calibrate the photometry, with precision of $\pm$ 0.2 mag over a century \citep{Grindlay2012}. This is being improved by also using the all-sky APASS CCD survey, which allows for improved photometry and color-corrections, improving the precision to $\pm$ 0.1 mag over a century \citep{Grindlay2012}. Currently these light curves are available for all sources with galactic latitudes greater than 30. Additionally, the DASCH team has released light curves for the Kepler field, calibrating this data using the \emph{Kepler} Input Catalog (KIC) \emph{g}-magnitudes. The light curves produced for the Kepler field have a precision of $\sim$ 0.1 mag \citep{Tang2013a}.

We have collected light curves from the DASCH website (\url{http://dasch.rc.fas.harvard.edu/}) and provide the specific details in replicating our light curve requests in Appendix \ref{sec:DASCHappendix} for replication purposes. To maintain high data quality, from every light curve we remove all points that a non-zero value for AFLAGS (defined by the DASCH interface as "'fatal' photometry pipeline errors"), and only analyze light curves that have at least 100 points after this cleaning of the data. We use light curves from both the APASS B-band and KIC \emph{g}-band calibrations.

\subsection{Selection of Comparison Stars}\label{sec:catalogs}
We selected two samples of comparison stars based on the properties of K8462842. From the Tycho-2 Spectral Type Catalog \citep{Wright2003}, we selected all F3 IV and V stars, as well as all F3 stars without an associated luminosity class. This was further cut to only include stars in DASCH's main data releases, corresponding to a galactic latitude of 30 degrees or larger. The resulting list has 971 stars for which we gathered light curves from the DASCH database.  These stars are predominantly between 9th and 11th magnitude and calibrated to the APASS B-band.

The second sample of comparison stars were chosen from the Kepler field based on the \emph{Kepler} Input Catalog \citep{Brown2011}. These stars were based on their similarities to K8462852, and represent all KIC stars within 100K of K8462852, as well as within 5\% in stellar radius and 10\% in log(g). The KIC lists K8462852 as having a temperature of 6584K, a log(g) of 4.124, and a stellar radius of 1.699 $R_{\astrosun}$. The resulting list had 559 stars that we gathered light curves for from the DASCH database, calibrated to the KIC \emph{g}-band.

\section{Results: Analysis of Long-term Photometric Trends}
\subsection{KIC 8462852}\label{sec:8462852}
\begin{figure}[!htb]
  \begin{center}
    \begin{subfigure}[b]{0.71\textwidth}
      \includegraphics[width=\textwidth]{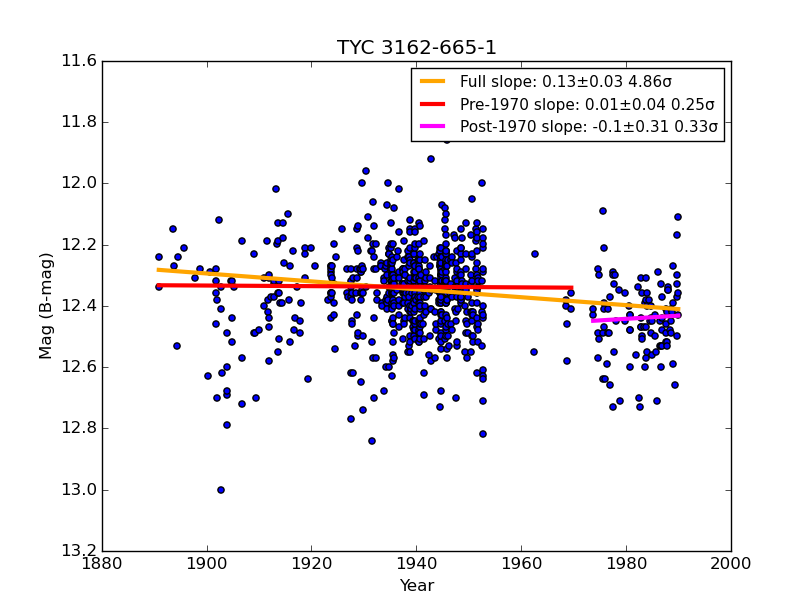}
    \end{subfigure}
    \begin{subfigure}[b]{0.71\textwidth}
      \includegraphics[width=\textwidth]{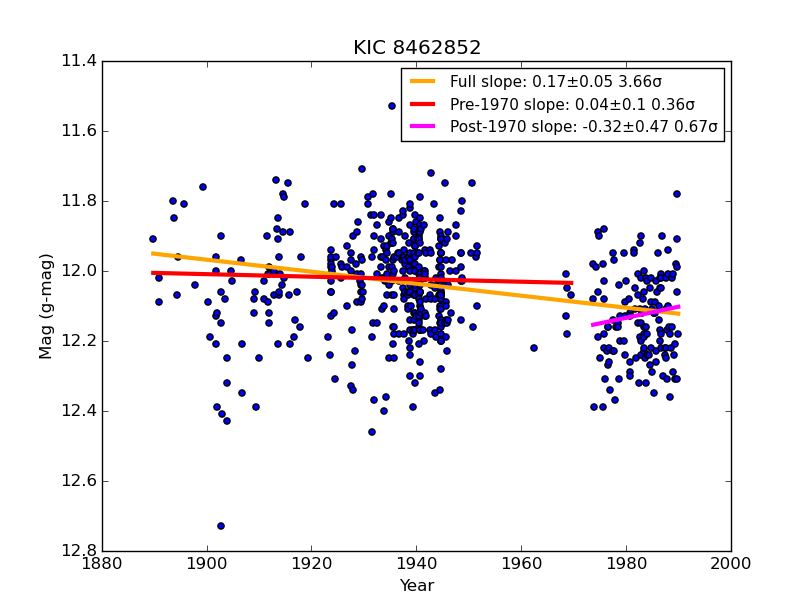}
    \end{subfigure}
  \end{center}
  \caption{KIC 8462852 light curves in the APASS B-band calibration (top) and KIC g-band calibration (bottom), with data cuts as described in \S\ref{sec:DASCH}. A linear fit to the entire light curve results in a dimming trend significant to at least 3 sigma. When the light curves are divided into two segments, before and after 1970, the linear fits for both segments are consistent with no slope.}
  \label{fig:KIC8462852}
\end{figure}

We first examine two light curves for KIC 8462852, one for each of the calibration catalogs that we use. The APASS B-band calibrated light curve consists of 1887 points, which is reduced to 781 points by our quality controls discussed in \S\ref{sec:DASCH}. The KIC g-band calibrated light curve consists of 1252 points, which is reduced to 575 points by our quality controls. In both cases, we use the method of weighted least squares to fit a single line to the full light curves, shown in Fig~\ref{fig:KIC8462852}. In both cases, we find that fitting a single line to the entire data results in a non-zero slope with at least 3 $\sigma$ significance, with the B-band slope of 0.129 $\pm$ 0.026 mag/century, and the g-band slope of 0.171 $\pm$ 0.047 mag/century.

We also break the light curves into two segments that we fit independently, before and after 1970; we explain our choice of this year in \S \ref{sec:twins}. When the light curves are broken at this year and using a critical p-value of 0.05, we cannot reject the hypothesis that the data is a flat line for any of the line segments. For both of the light curves we have also used a Two Sample Weighted T-Test to test if the magnitudes before and after 1970 are drawn from the same population \citep{Welch1947}. We are able to reject the hypothesis that pre-1970 and post-1970 data are drawn from the same distribution with a critical p-value of 0.05. These results suggest that the century-long slope we observe is a result of an abrupt change in the magnitude around 1970 rather than a gradual and consistent dimming over the course of a century.

\subsection{Comparison Stars}\label{sec:twins}

\begin{figure}[!htb]
  \begin{center}
   \includegraphics[width=\textwidth]{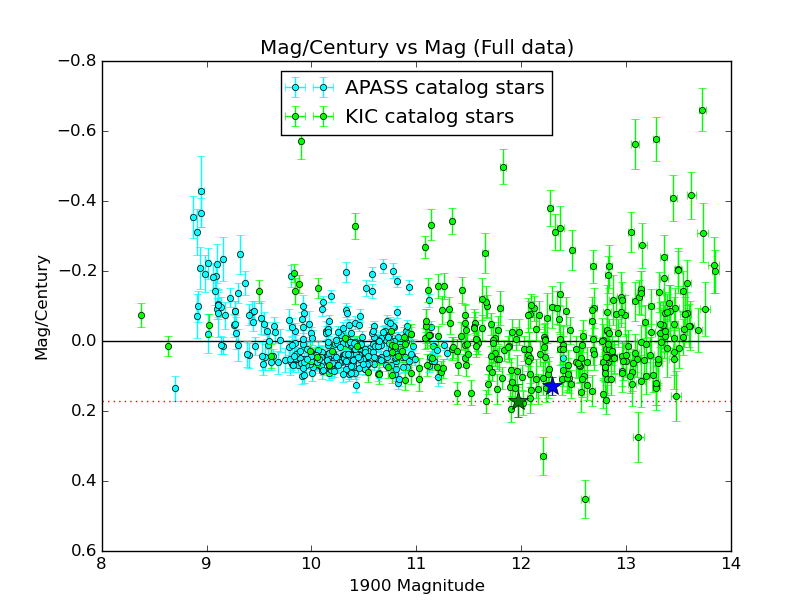}
  \end{center}
  \caption{Century-long slopes as a function of magnitude for all stellar twins of KIC 8462852. Cyan points are stars from the Tycho catalog calibrated to the APASS catalog B-band. Lime points are stars from the KIC and calibrated to the KIC g-band. The dark blue and dark green star symbols are KIC 8462852 calibrated to the APASS B-band and KIC g-band, respectively. The red dotted line is our calculated slope for KIC 8462852 from the KIC g-band.}
  \label{fig:magc_vs_mag}
\end{figure}

In addition to looking at KIC 8462852 individually, we also compare it to a large number of comparison stars. We start with the two lists of stars we discussed in \S \ref{sec:catalogs}. We then apply the same cuts for quality control, and are left with 337 stars from the Tycho-2 Spectral Type Catalog and 307 stars from the KIC with at least 100 data points. We calculate the slope of a simple line using the method of weighted least squares for the light curves of all 644 stars, as shown in Fig~\ref{fig:magc_vs_mag}. Since there is no astrophysical reason to expect that long-term brightness changes would be present as a function of magnitude, the trends towards brightening that we see for both bright ($\leq$ 9) stars and faint ($\geq$ 13.5) stars are likely indicative of systematics at the bright and faint limits of the calibration, respectively. Within these magnitude limits, KIC 8462852 is not unique in having a significant slope, and a large number of stars show slopes that deviate from a flat slope by at least 1 $\sigma$. Even at approximately the same magnitude, there are several stars with more significant slopes, either dimming or brightening.

\begin{figure}[!htb]
  \begin{center}
    \begin{subfigure}[b]{0.6\textwidth}
      \includegraphics[width=\textwidth]{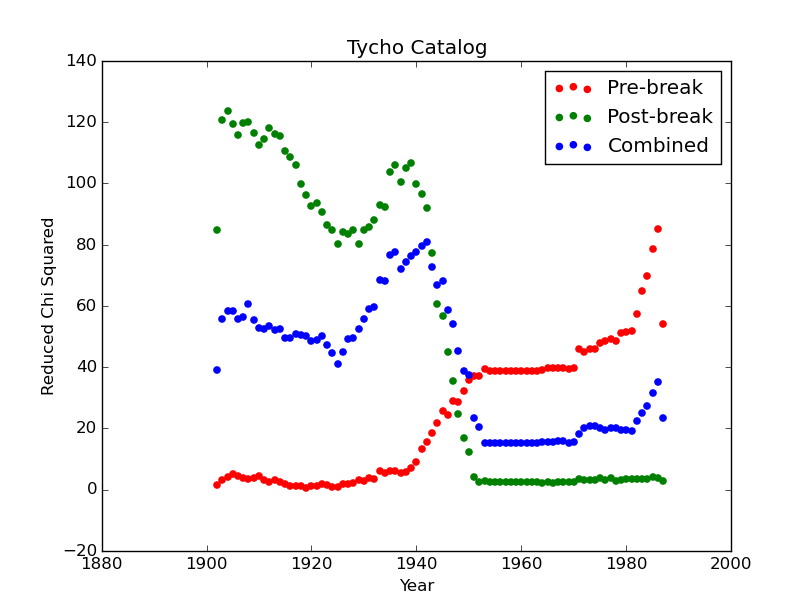}
    \end{subfigure}
    \begin{subfigure}[b]{0.6\textwidth}
      \includegraphics[width=\textwidth]{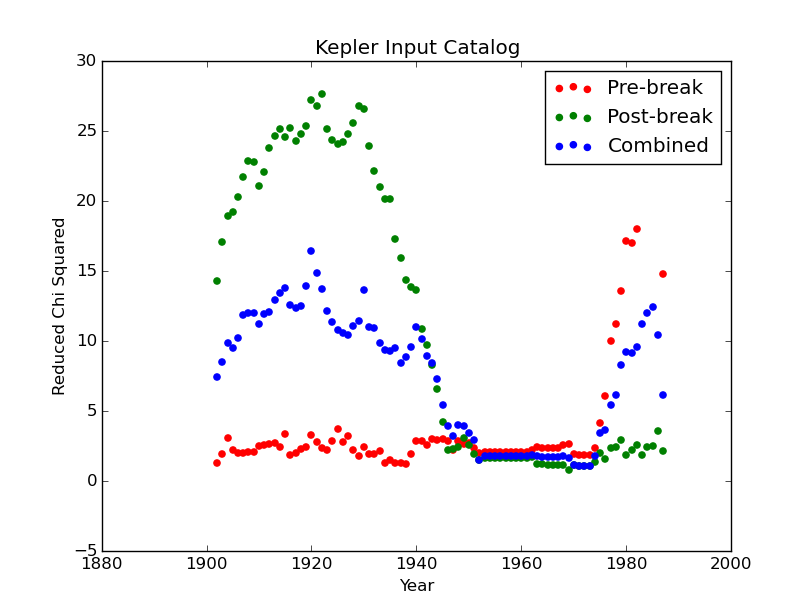}
    \end{subfigure}
  \end{center}
  \caption{Reduced $\chi^{2}$ values of the t-statistic of the slope as compared to a normal distribution and as a function of the year at which the light curves are split. The red points are the comparison to a normal distribution only for slopes prior to the break, the green points are only slopes after the break, and the blue points are the reduced $\chi^2$ for the slopes before and after the break jointly. The Tycho catalog (top) reaches a minimum in 1954, and the KIC (bottom) reaches a minimum in 1971. The Menzel gap is generally defined as being from 1953 to 1969.}
  \label{fig:chisq}
\end{figure}

We next propose a simple hypothesis to look for a systematic shift in the data. If we presume that the light curses of all stars in the sample are on average flat over a century of observations, then we should expect that the t-statistics for the slopes follow a normal distribution In other words, we expect that just by the nature of noise in the data, we would expect to find $\sim$ 200 stars with slopes greater than 1 $\sigma$ from 0, $\sim$ 30 stars with slopes greater than 2 $\sigma$ from 0, and only 1 or 2 stars with slopes greater than 3 $\sigma$ from 0. This makes the slope distribution found in \S \ref{sec:twins} seem far more unlikely than it appears to be in Fig~\ref{fig:magc_vs_mag}. We then add to our hypothesis the possibility of a systematic structural break in the data, such that at some year the brightness for all the stars is shifted. To determine the best year to place this break, we iterate over 1902 to 1987; for each year, we split every light curve at that year, fit each of the segments to a line, and calculate the t-statistics for the slopes. We then use a Pearson's $\chi^{2}$ test to compare the distribution of the t-statistics of the slopes to a normal distribution with the intention of finding the year where the t-statistics are closest to following a normal distribution. We then calculate the reduced $\chi^{2}$ for this distribution, shown in Fig~\ref{fig:chisq} and treating the Tycho and KIC stellar samples independently. We find that the optimal break using the light curves based on the Tycho catalog calibrations is 1954, and the optimal break using the light curves based on the for the KIC stars is 1971. More significantly, however, we see that there is a region in both figures where the reduced $\chi^{2}$ is approximately the same, and this roughly corresponds to the Menzel gap, which spans 1953 to 1969 \citep{Schaefer2016}. As the reduced $\chi^{2}$ is approximately the same over this time frame, we use 1970 as a break year going forward for both sets of stars.

\begin{figure}[!htb]
  \begin{center}
    \begin{subfigure}[b]{0.6\textwidth}
      \includegraphics[width=\textwidth]{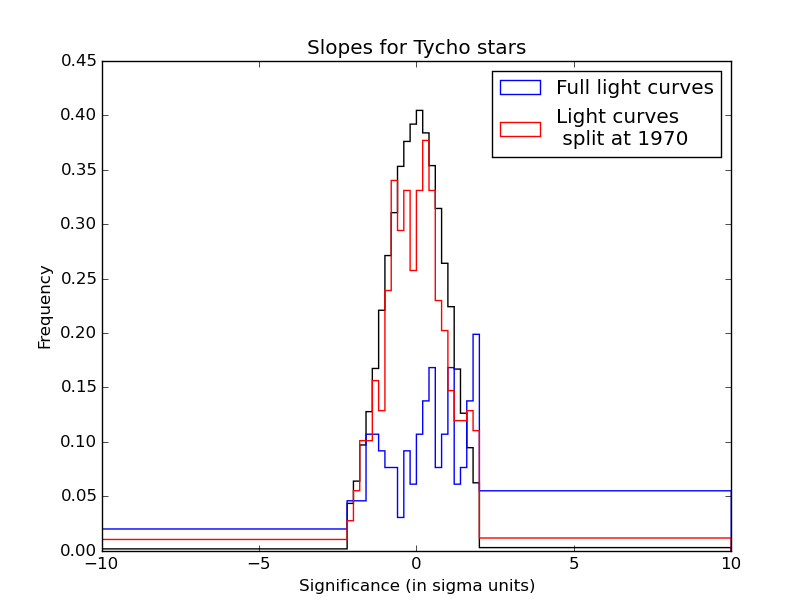}
    \end{subfigure}
    \begin{subfigure}[b]{0.6\textwidth}
      \includegraphics[width=\textwidth]{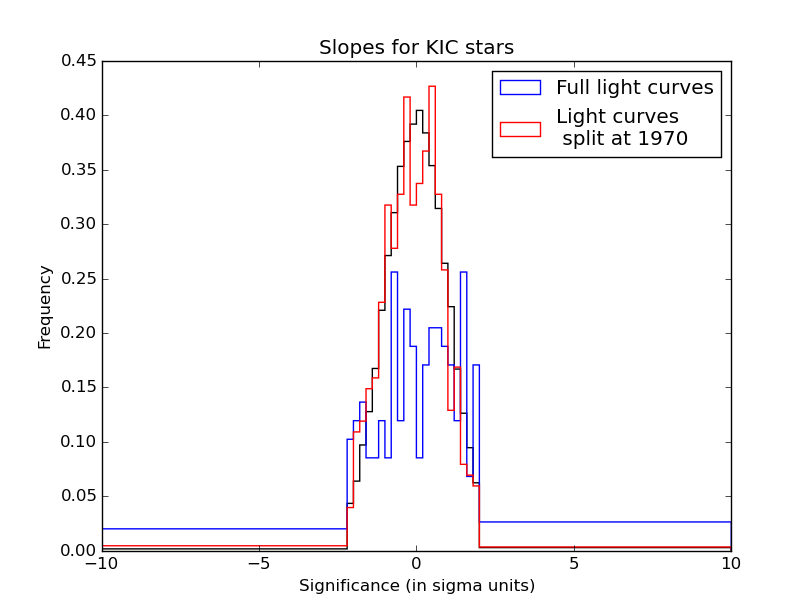}
    \end{subfigure}
  \end{center}
  \caption{Histograms of the t-statistic for the slopes of all stars in the Tycho catalog (top) and KIC (bottom). The black line marks the histogram of a normal distribution. The blue line is the distribution of t-statistics of the slopes when we fit the entire light curve at once. The red line is the distribution of the t-statistics of the slopes when we split each light curve in two pieces at 1970 before fitting lines to the light curves.}
  \label{fig:hist}
\end{figure}

In Fig~\ref{fig:hist} we compare the distributions of the t-statistics of the slopes. It becomes readily apparent that by splitting the light curves at 1970 and fitting the segments before and after this year independently, we have a distribution that is much more consistent with all the stars having slopes of 0 and that the presence of slopes for some of the stars is a result of noise.

\begin{figure}[!htb]
  \begin{center}
   \includegraphics[width=\textwidth]{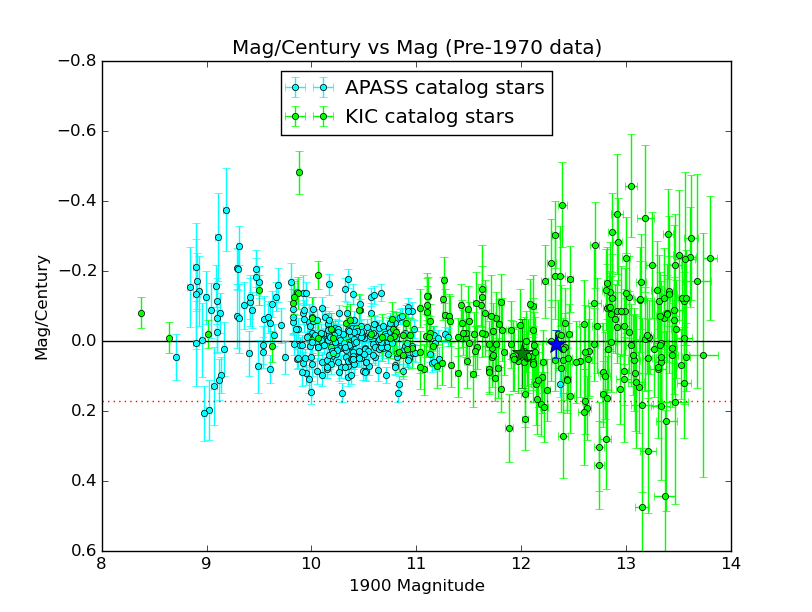}
  \end{center}
  \caption{Slopes as a function of magnitude for all stellar twins of KIC 8462852, using only the pre-1970 light curves. Cyan points are stars from the Tycho catalog calibrated to the APASS catalog B-band. Lime points are stars from the KIC and calibrated to the KIC g-band. The dark blue and dark green stars are KIC 8462852 calibrated to the APASS B-band and KIC g-band, respectively. The red dotted line is our calculated slope for KIC 8462852 from the KIC \emph{g}-band and without excluding post-1970 data.}
  \label{fig:magc_vs_mag_pre1970}
\end{figure}

We reassess the distribution of slopes as a function of magnitude in Fig~\ref{fig:magc_vs_mag_pre1970}, where we have now only used the pre-1970 data in the weighted least-squares linear fit. The errors on each point are slightly larger, but the brightening trends seen for bright and faint stars is less noticeable as those points are closer to being centered on 0. Additionally, the values for KIC 8462852 are now consistent with a slope of 0 (the slopes for the post-1970 data in KIC 8462852 is also displayed in Fig~\ref{fig:KIC8462852}).

\section{Discussion}
The DASCH light curves for KIC 8462852, when examined on their own and with a single linear fit, suggest that the star has undergone a century-long dimming event that would be particularly noteworthy given the events in the \emph{Kepler} light curve that were announced by \citet{Boyajian2016}. The KIC 8462852 light curve, however, can also be explained by a structural break in an otherwise flat light curve. When all light curves are split at 1970, we find that the t-statistics of the slopes are much closer to a normal distribution, as would be expected if all light curves were truly flat. We do still find an excess of stars with significant slopes, with 9 stars from the APASS B-band calibration and 2 stars from the KIC g-band calibration having slopes that are non-zero to greater than 5 $\sigma$. All of these stars are listed in Table~\ref{table:stars}, and the light curves are included in Appendix~\ref{sec:Targets}. These stars can be accounted for with three possible explanations: (1) there are additional structural breaks beyond the Menzel Gap that have not been accounted for (2) systematics such as slightly incorrect error estimations result in underestimating the error in the slope (3) some of these stars may actually have changed brightness over the last century. While these stars are most likely indications of additional systematics, it is possible that one or more of them may represent an astrophysical signal.

\begin{table}[!htb]
\centering
\caption{Stars with Significant ($5 \sigma$) Long-term Photometric Trends}
\label{table:stars}
\begin{tabular}{lrrrl}
Star   & Full Slope & Pre-1970 Slope      & Post-1970 Slope \\
 TYC-6727-00524-1 &   0.121$\pm$0.017 &   0.150$\pm$0.025 &   0.076$\pm$0.258 \\
 TYC-6174-00949-1 &  -0.215$\pm$0.020 &  -0.138$\pm$0.024 &  -0.264$\pm$0.340 \\
 TYC-5531-01038-1 &  -0.169$\pm$0.016 &  -0.144$\pm$0.021 &  -0.321$\pm$0.270 \\
 TYC-5554-01593-1 &  -0.023$\pm$0.019 &  -0.161$\pm$0.031 &   0.120$\pm$0.263 \\
 TYC-6178-00821-1 &  -0.088$\pm$0.020 &  -0.178$\pm$0.028 &  -0.213$\pm$0.277 \\
 TYC-6749-00508-1 &   0.126$\pm$0.021 &   0.141$\pm$0.025 &   0.108$\pm$0.411 \\
 TYC-6160-00274-1 &   0.111$\pm$0.017 &   0.124$\pm$0.022 &   0.158$\pm$0.325 \\
 TYC-5554-01017-1 &  -0.022$\pm$0.017 &  -0.130$\pm$0.026 &   0.603$\pm$0.275 \\
 TYC-6165-01434-1 &   0.096$\pm$0.021 &   0.149$\pm$0.027 &  -0.198$\pm$0.409 \\
         K3868420 &  -0.571$\pm$0.051 &  -0.481$\pm$0.061 &   1.101$\pm$1.422 \\
        K11802860 &  0.452$\pm$0.055 &   1.090$\pm$0.180 &  -1.282$\pm$0.487
\end{tabular}
\end{table}

\section{Conclusion}
The DASCH survey is extremely useful for looking for large changes in flux that have taken place with a century-baseline for a range of sources, including long-period eclipsing binaries \citep{Rodriguez2016}, large-amplitude variation in K giants \citep{Tang2010}, and quasars/blazars \citep{Grindlay2012}. There are, however, limitations with the lowest errors still being on the order of 0.1 mag, and this results in circumstances that allow for systematics to be mistaken for astrophysical signals. We believe that KIC 8462852 provides an example of this, and that slight a slight systematic shift during the Menzel Gap is the cause for the slope that is found with a simple linear fit to the light curves, and in all probability, KIC 8462852 itself has not undergone any long-term trends in brightness. We do, however, present a list of several stars that still show significant slopes that may be worth further investigation to either better characterize the systematics in the DASCH data for small magnitude changes, or to determine if these stars are undergoing significant and unexpected changes in magnitude on long time scales.

\acknowledgments
\emph{Acknowledgments.} The authors would like to thank Kyle Conroy for multiple helpful discussions on this research.
The DASCH project at Harvard is grateful for partial support from NSF grants AST-0407380, AST-0909073, and AST-1313370.
%
%
%

% KIC 8462852
% Slope: 0.171525234249 ± 0.0465759083515 Mag/century p-value 0.000252616535334 z-score 3.6595926114
% TYC 3162-665-1
% Slope: 0.129309402893 ± 0.0264003750119 Mag/century p-value 1.17763522047e-06 z-score 4.85936386993

%\subsection{DASCH}
%\subsection{Light Curve Reduction}
%Tycho
%Combined
%[1954 1956 1958 1955 1957 1959 1960 1961 1962 1969]
%[ 15.10688766  15.14420717  15.14420717  15.14420717  15.14420717
%  15.15428165  15.15428165  15.15428165  15.15428165  15.19804154]
%KIC minimum years
%Combined
%[1971 1972 1973 1970 1952 1969 1964 1967 1966 1965]
%[ 1.06527146  1.06527146  1.06527146  1.10792146  1.47045606  1.59738102
%  1.71413621  1.71884639  1.71884639  1.71884639]

\bibliographystyle{apalike}
\bibliography{libAAS}

\begin{thebibliography}{}

\bibitem[Bodman and Quillen, 2016]{Bodman2016}
Bodman, E. H.~L. and Quillen, A. (2016).
\newblock {Kic 8462852: Transit of a Large Comet Family}.
\newblock {\em \apj}, 819(2):L34.

\bibitem[Boyajian et~al., 2016]{Boyajian2016}
Boyajian, T.~S., LaCourse, D.~M., Rappaport, S.~A., et~al. (2016).
\newblock {Planet Hunters IX. KIC 8462852 – where's the flux?}
\newblock {\em \mnras}, 457(4):3988--4004.

\bibitem[Brown et~al., 2011]{Brown2011}
Brown, T.~M., Latham, D.~W., Everett, M.~E., et~al. (2011).
\newblock {{\textless}i{\textgreater}KEPLER{\textless}/i{\textgreater} INPUT
  CATALOG: PHOTOMETRIC CALIBRATION AND STELLAR CLASSIFICATION}.
\newblock {\em \aj}, 142(4):112.

\bibitem[Grindlay et~al., 2012]{Grindlay2012}
Grindlay, J., Tang, S., Los, E., et~al. (2012).
\newblock {Opening the 100-Year Window for Time-Domain Astronomy}.
\newblock {\em Proceedings of the International Astronomical Union},
  7(S285):29--34.

\bibitem[Harp et~al., 2015]{Harp2015}
Harp, G.~R., Richards, J., Shostak, S., et~al. (2015).
\newblock {Radio SETI Observations of the Anomalous Star KIC 8462852}.

\bibitem[Hippke et~al., 2016]{Hippke2016a}
Hippke, M., Angerhausen, D., Lund, M.~B., et~al. (2016).
\newblock {KIC 8462852 did likely not fade during the last 100 years}.

\bibitem[Laycock et~al., 2010]{Laycock2010}
Laycock, S., Tang, S., Grindlay, J., et~al. (2010).
\newblock {Digital Access To a Sky Century At Harvard: Initial Photometry and
  Astrometry}.
\newblock {\em \aj}, 140(4):1062--1077.

\bibitem[Marengo et~al., 2015]{Marengo2015}
Marengo, M., Hulsebus, A., and Willis, S. (2015).
\newblock {Kic 8462852: the Infrared Flux}.
\newblock {\em \apj}, 814(1):L15.

\bibitem[Rodriguez et~al., 2016]{Rodriguez2016}
Rodriguez, J.~E., Stassun, K.~G., Lund, M.~B., et~al. (2016).
\newblock {An Extreme Analogue of Epsilon Aurigae: An M-giant
  Eclipsed Every 69 Years by a Large Opaque Disk Surrounding a Small Hot
  Source}.
\newblock (2014).

\bibitem[Schaefer, 2016]{Schaefer2016}
Schaefer, B.~E. (2016).
\newblock {KIC 8462852 Faded at an Average Rate of 0.165+-0.013 Magnitudes Per
  Century From 1890 To 1989}.
\newblock pages 1--11.

\bibitem[Schuetz et~al., 2015]{Schuetz2015}
Schuetz, M., Vakoch, D.~A., Shostak, S., et~al. (2015).
\newblock {Optical SETI Observations of the Anomalous Star KIC 8462852}.

\bibitem[Tang et~al., 2010]{Tang2010}
Tang, S., Grindlay, J., Los, E., et~al. (2010).
\newblock {Dasch Discovery of Large Amplitude ∼10-100 Year Variability in K
  Giants}.
\newblock {\em \apj}, 710(1):L77--L81.

\bibitem[Tang et~al., 2013]{Tang2013a}
Tang, S., Grindlay, J., Los, E., et~al. (2013).
\newblock {Improved Photometry for the DASCH Pipeline}.
\newblock {\em \pasp}, 125(929):857--865.

\bibitem[Thompson et~al., 2016]{Thompson2015c}
Thompson, M.~A., Scicluna, P., Kemper, F., et~al. (2016).
\newblock {Constraints on the circumstellar dust around KIC 8462852}.
\newblock {\em \mnras}, 458(1):L39.

\bibitem[Welch, 1947]{Welch1947}
Welch, B.~L. (1947).
\newblock {THE GENERALIZATION OF ‘STUDENT'S' PROBLEM WHEN SEVERAL DIFFERENT
  POPULATION VARLANCES ARE INVOLVED}.
\newblock {\em Biometrika}, 34(1-2):28--35.

\bibitem[Wright et~al., 2003]{Wright2003}
Wright, C.~O., Egan, M.~P., Kraemer, K.~E., et~al. (2003).
\newblock {The Tycho-2 Spectral Type Catalog}.
\newblock {\em \aj}, 125(1):359.

\bibitem[Wright et~al., 2015]{Wright2015a}
Wright, J.~T., Cartier, K. M.~S., Zhao, M., et~al. (2015).
\newblock {the Ĝ Search for Extraterrestrial Civilizations With Large Energy
  Supplies. Iv. the Signatures and Information Content of Transiting
  Megastructures}.
\newblock {\em \apj}, 816(1):17.

\end{thebibliography}

\appendix
\section{DASCH Light Curve Access}\label{sec:DASCHappendix}
Different formats for obtaining light curves from the DASCH web site can result in variations in the light curve that is returned, and so we outline the procedure for obtaining the light curves here. The DASCH light curves are accessible through \url{http://dasch.rc.fas.harvard.edu/lightcurve.php}. Stars were queried in batches of 10, using the default settings for minimum number of measurements (1), distance in arcseconds (5), and using a frame format. From the upper-left frame of the next window, we then selected "Download all points in table form." This provides a list of the results for all sources in the query. The downloaded files are the short form Starbase (tab-delimited ASCII) tables, also listed as option 'A' for each source.

\section{Stars with Significant Slopes}\label{sec:Targets}
\begin{figure}[!htb]
  \begin{center}
    \begin{subfigure}[b]{\textwidth}
      \includegraphics[width=0.49\textwidth]{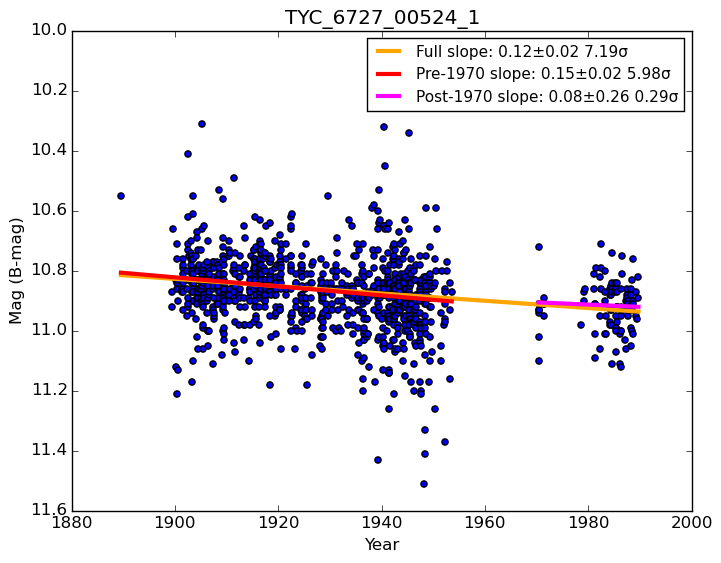}
      \includegraphics[width=0.49\textwidth]{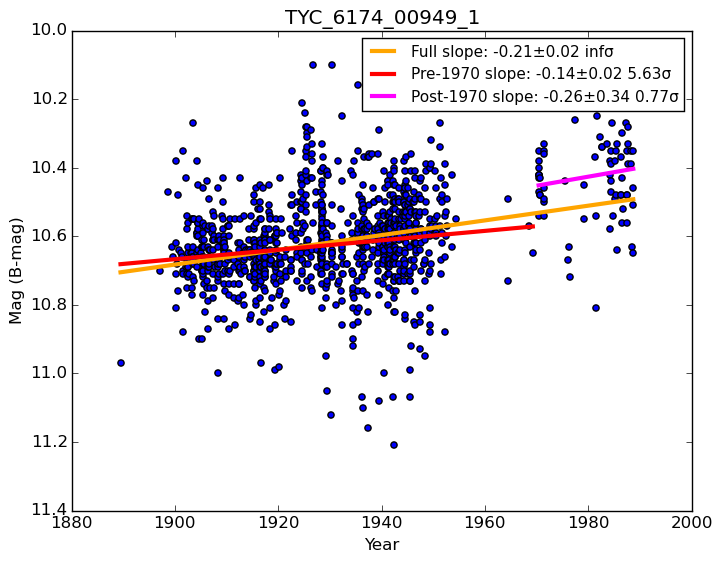}
    \end{subfigure}
    \begin{subfigure}[b]{\textwidth}
      \includegraphics[width=0.49\textwidth]{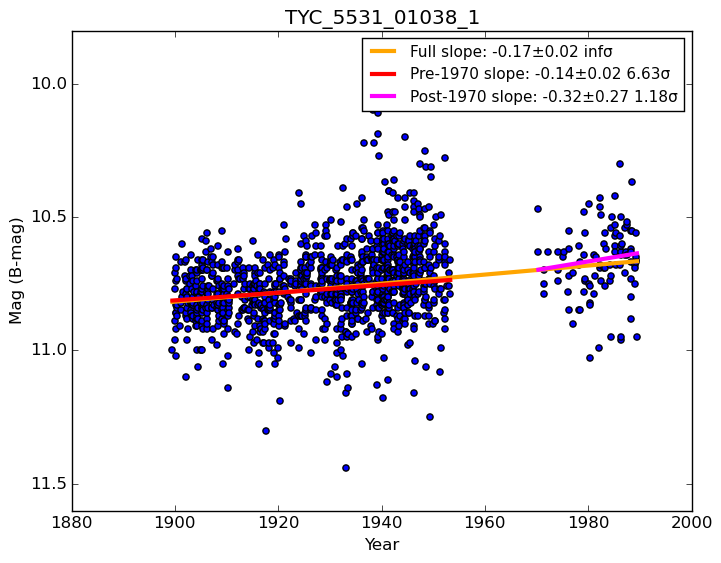}
      \includegraphics[width=0.49\textwidth]{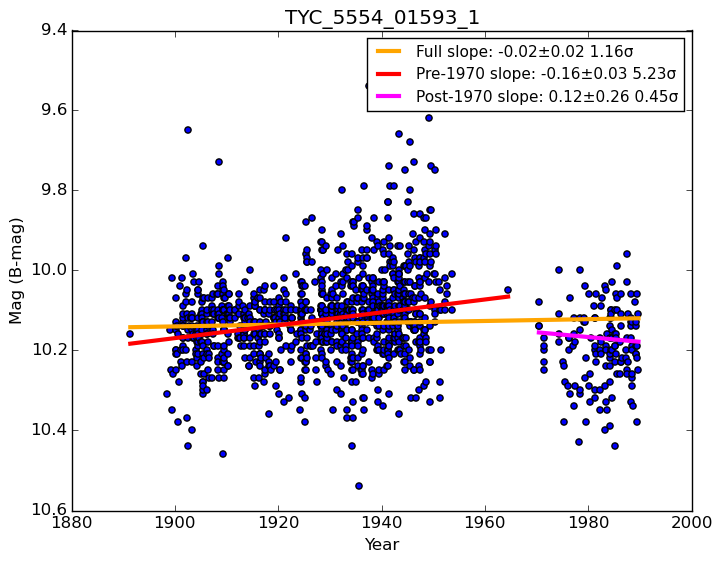}
    \end{subfigure}
    \begin{subfigure}[b]{\textwidth}
      \includegraphics[width=0.49\textwidth]{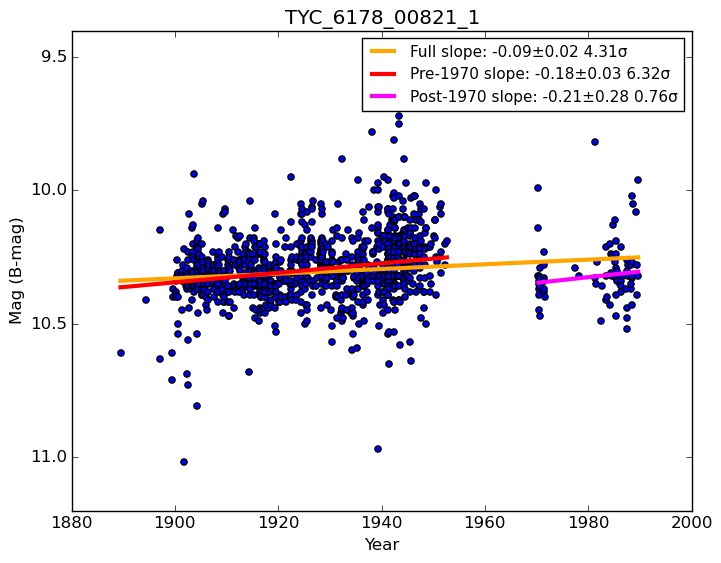}
      \includegraphics[width=0.49\textwidth]{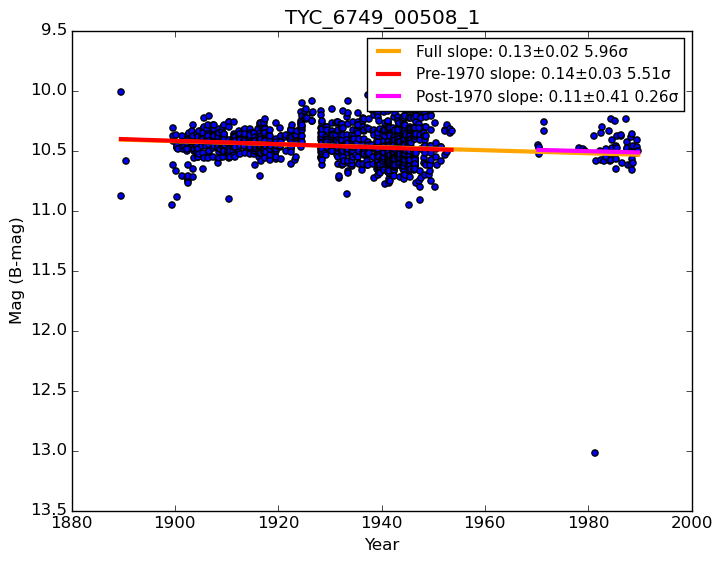}
    \end{subfigure}
  \end{center}
  \caption{Stars showing slopes greater than 5 $\sigma$ from 0 either pre-1970 or post-1970.}
  \label{fig:Extra_1}
\end{figure}

\begin{figure}[!htb]
  \begin{center}
    \begin{subfigure}[b]{\textwidth}
      \includegraphics[width=0.49\textwidth]{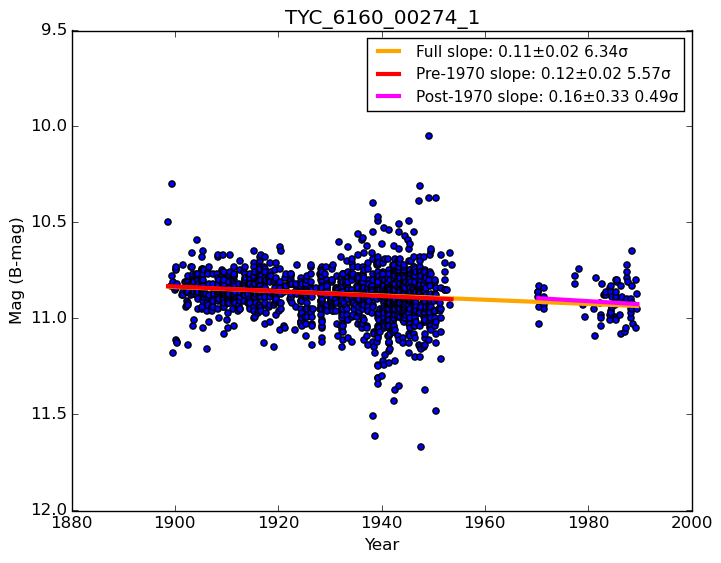}
      \includegraphics[width=0.49\textwidth]{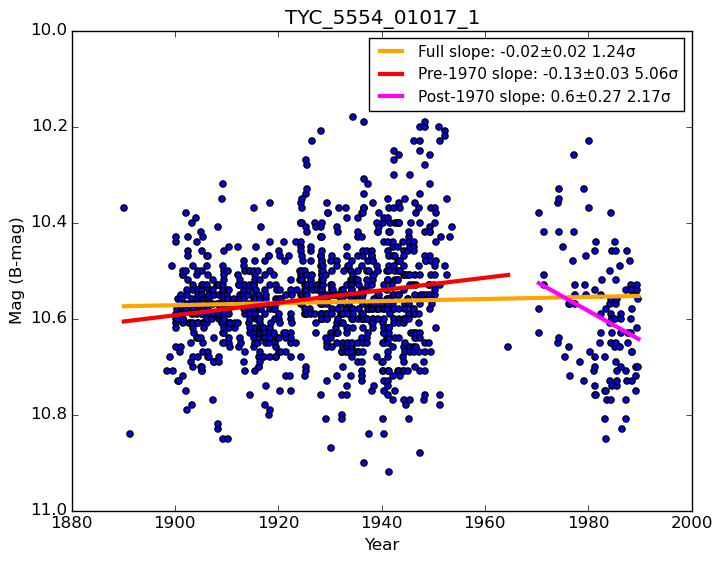}
    \end{subfigure}
    \begin{subfigure}[b]{\textwidth}
      \includegraphics[width=0.49\textwidth]{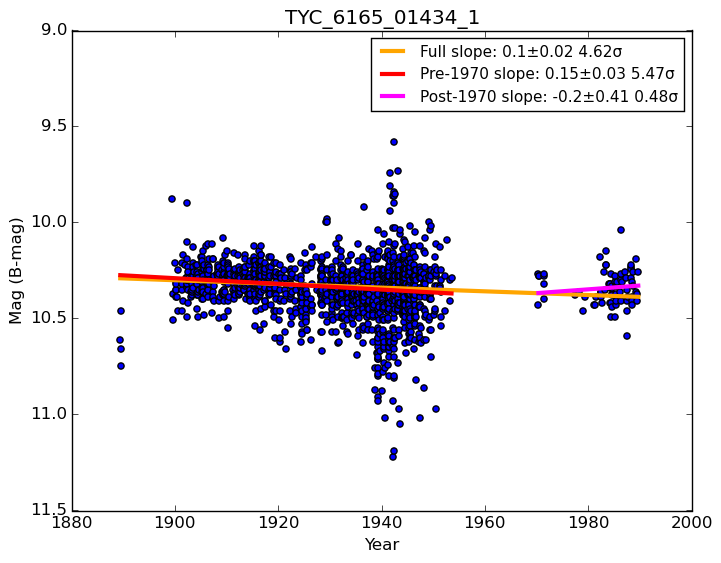}
    \end{subfigure}
    \begin{subfigure}[b]{\textwidth}
      \includegraphics[width=0.49\textwidth]{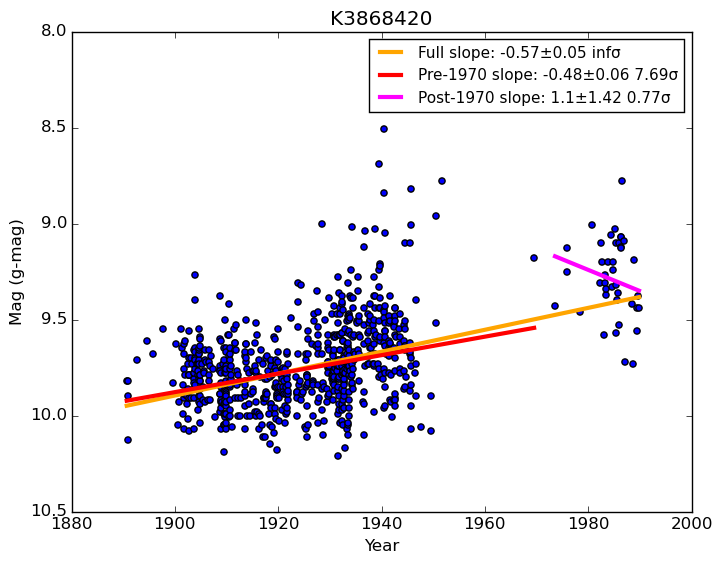}
      \includegraphics[width=0.49\textwidth]{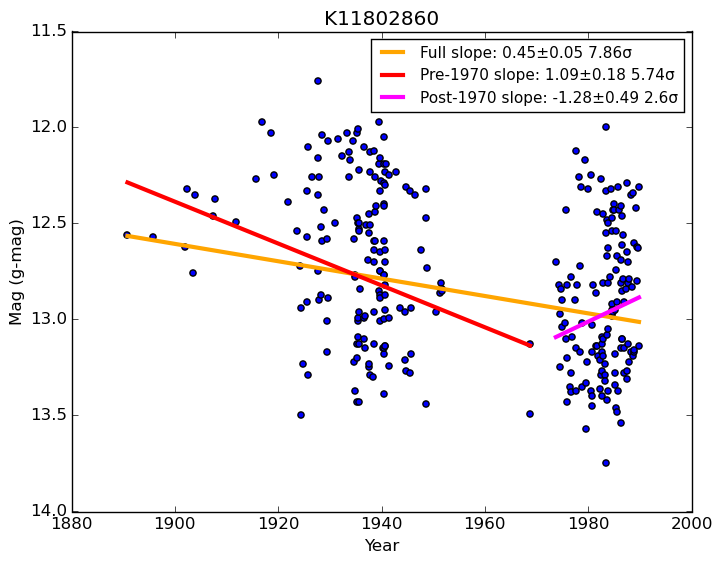}
    \end{subfigure}
  \end{center}
  \caption{Stars showing slopes greater than 5 $\sigma$ from 0 either pre-1970 or post-1970.}
  \label{fig:Extra_2}
\end{figure}

In the analysis in this paper, we examined over 600 stars, and we fit over 1200 light curves after the data is split at 1970. If these slopes are purely a result of statistical noise in flat light curves, the chance of having even one slope that is significant by more than 5 $\sigma$ is extremely small. Instead, we find 11 light curves that exceed this threshold. These stars may simply be indicative of further systematics that we have not addressed, but we present them here in Figures~\ref{fig:Extra_1} and \ref{fig:Extra_2} in the case that these may also represent astrophysical behavior.

\end{document}